\documentclass[a4paper,11pt]{article}

\usepackage[dvips]{graphicx,psfrag}
\usepackage{amssymb}

\makeatletter

 \@addtoreset{equation}{section}
\makeatother
\newcommand{\beq}{\begin{eqnarray}}
\newcommand{\eeq}{\end{eqnarray}}
\def\tr{\mathop{\mathrm{tr}}\nolimits}

\begin{document}

\begin{titlepage}

\begin{flushright}
OU-HET 430\\
{\tt hep-th/0302090}\\
Feb 2003
\end{flushright}
\bigskip

\begin{center}
{\LARGE\bf
A New Class of Conformal Field Theories with Anomalous Dimensions 
}
\vspace{1cm}

\setcounter{footnote}{0}
\bigskip

\bigskip
{\renewcommand{\thefootnote}{\fnsymbol{footnote}}
{\large\bf Kiyoshi Higashijima\footnote{
     E-mail: {\tt higashij@phys.sci.osaka-u.ac.jp}} and
 Etsuko Itou\footnote{
     E-mail: {\tt itou@het.phys.sci.osaka-u.ac.jp}}
}}

\vspace{4mm}

{\sl
Department of Physics,
Graduate School of Science, Osaka University,\\ 
Toyonaka, Osaka 560-0043, Japan \\
}

\end{center}
\bigskip

\begin{abstract}
The Wilsonian renormalization group (WRG) equation is used to derive 
a new class of scale invariant field theories with nonvanishing 
anomalous dimensions in $2$-dimensional ${\cal N}=2$ supersymmetric 
nonlinear sigma models. When the coordinates of the target manifolds 
have nontrivial anomalous dimensions, vanishing of the $\beta$ function 
suggest the existence of novel conformal field theories whose target 
space is not Ricci flat.
We construct such conformal field theories with ${\bf U}(N)$ symmetry. 
The theory has one free parameter $a$ corresponding to the anomalous 
dimension of the scalar fields. The new conformal field theories are 
well behaved for positive $a$, while they have curvature singularities 
at the boundary for $a<0$. When the target space is of complex 
$1$-dimension, we obtain the explicit form of the Lagrangian, which 
reduces to two different kinds of free field theories in weak and 
in strong coupling limit. The target space in this case looks 
like a semi-infinite cigar with one-dimension compactified to a circle. 

\end{abstract}

\end{titlepage}

\pagestyle{plain}

\section{Introduction}
Nonlinear sigma models (NL$\sigma$Ms) in two dimensions are interesting 
for several reasons.
They help us to understand various non-perturbative phenomena in four 
dimensional gauge theories such as confinement or dynamical mass generation 
\cite{HKNT,Morozov}. They also provide the description of superstrings 
propagating in the curved space-time. In the latter case, the consistency of 
strings requires the  ${\cal N}=2$ superconformal symmetry of the NL$\sigma$Ms. 
Since ${\cal N}=2$ supersymmetry and the scale invariance imply ${\cal N}=2$ 
superconformal symmetry, these NL$\sigma$Ms have to be scale invariant. 
In quantum field theories, the scale invariance, suffering from anomaly 
due to the divergent renormalization effects, is realized only at the 
fixed-points of the renormalization group equation. Since field theories 
at the fixed-point also describes the phase transition, it is important 
to study these fixed-point theories of ${\cal N}=2$ supersymmetric 
NL$\sigma$Ms. 

In ${\cal N}=2$ supersymmetric NL$\sigma$Ms, the field variables take 
values on the complex curved spaces called K\"{a}hler manifolds, 
whose metrics are specified completely by the K\"{a}hler potentials.
These K\"{a}hler potentials, arbitrary functions of the chiral superfields, 
have infinite numbers of coupling constants since any NL$\sigma$M is 
renormalizable in perturbation theories in two dimensions. It is convenient 
to use the Wilsonian renormalization group (WRG) equation for the 
nonperturbative study of field theories with infinite numbers of 
coupling constants. 
In a previous paper, we derived the $\beta$ function for 
$2$-dimensional ${\cal N}=2$ supersymmetric NL$\sigma$M using the 
WRG equation\cite{HI,CHL,CL}:  
\beq
\beta (g_{i \bar{j}})=\frac{1}{2\pi} R_{i \bar{j}} +\gamma \Big[\varphi^k g_{i \bar{j},k}+\varphi^{* \bar{k}}g_{i \bar{j}, \bar{k}}+2 g_{i \bar{j}} \Big].
\eeq
The WRG equation shows the variation of Wilsonian effective action 
when the cutoff scale is changed 
\cite{Wilson Kogut,Wegner and Houghton,Morris,Aoki}. 
The first term, proportional to 
the Ricci tensor of the target space, comes from the one-loop diagrams, 
whereas the second term, proportional to the anomalous dimension 
$\gamma$ of fields, comes from the rescaling of fields to normalize 
the kinetic term properly. The presence of the anomalous dimension 
reflects the nontrivial continuum limit of the fields.

When the anomalous dimension of the field vanishes, the 
scale invariance is realized for NL$\sigma$Ms on the Ricci-flat 
K\"{a}hler (Calabi-Yau) manifolds \cite{AFM}. Calabi-Yau metrics 
for some noncompact manifolds have been explicitly constructed 
\cite{HKN}, when the number of isometries is sufficient to reduce 
the Einstein equation to an ordinary differential equation. 

On the other hand, when the anomalous dimension of fields does not 
vanish, the condition of the scale invariance is 
quite different.  In this article, we study the novel conformal 
field theories with anomalous dimension by solving the condition 
of the fixed-point: $\beta =0$. We will assume ${\bf U}(N)$ symmetry 
to reduce a set of the partial differential equations to an ordinary 
differential equation. The conformal theories obtained have one free 
parameter corresponding to the anomalous dimension of the scalar fields.
The geometry of the target manifolds crucially depends on the sign of 
the anomalous dimensions.

This paper is organized as follow:
In \S \ref{review}, we review briefly the WRG equation for $2$ dimensional ${\cal N}=2$ supersymmetric nonlinear sigma model.
In \S \ref{solution}, we derive the condition of scale invariance assuming 
the ${\bf U}(N)$ symmetry.
In \S \ref{geometry}, we study the geometry of target spaces of conformal field theories.
In \S \ref{fixedpoint}, the properties of the new conformal field theories are discussed.

\section{Wilsonian Renormalization Group equation}\label{review}
In this section, let us recapitulate the Wilsonian renormalization group equation for the ${\cal N}=2$ supersymmetric NL$\sigma$M.
The Wilsonian renormalization group equation describes the variation of effective action when the cutoff energy scale $\Lambda$ is changed to $\Lambda (\delta t)=\Lambda e^{-\delta t}$ in $D$ dimensional field theory \cite{Wilson Kogut, Wegner and Houghton, Morris, Aoki}:
\beq  
\frac{d}{dt}S[\Omega; t]&=&\frac{1}{2\delta t} \int_{p'} \tr \ln \left(\frac{\delta^2 S}{\delta \Omega^i \delta \Omega^j}\right)\nonumber\\
&&-\frac{1}{2 \delta t}\int_{p'} \int_{q'} \frac{\delta S}{\delta \Omega^i (p')} \left(\frac{\delta^2 S}{\delta \Omega^i (p')\delta \Omega^j (q')} \right)^{-1} \frac{\delta S}{\delta \Omega^j (q')} \nonumber\\
&&+ \left[D-\sum_{\Omega_i} \int_p \hat{\Omega}_i (p) \left(d_{\Omega_i}+\gamma_{\Omega_i}+\hat{p}^{\mu} \frac{\partial}{\partial \hat{p}^{\mu}} \right) \frac{\delta}{\delta \hat{\Omega}_i (p)} \right] \hat{S},\nonumber\\ \label{WRG-1} 
\eeq
where $d_{\Omega}$ and $\gamma_{\Omega}$ denote the canonical and anomalous dimensions of the field $\Omega$.
The caret indicates dimensionless quantities.
The first and second terms in eq.(\ref{WRG-1}) correspond to the one-loop and tree diagrams where internal lines are eliminated fields with high momentum $\Lambda (\delta t)<p<\Lambda$. The remaining terms come from the rescaling of fields to normalize the coefficient of the kinetic term to unity.

This WRG equation is an infinite set of differential equations for various coupling constants in the most general action $S$.
In practice, we usually expand the action in power of derivatives and retain the first few terms.
We often introduce symmetry to further reduce the number of independent coupling constants.

We impose ${\cal N}= 2$ supersymmetry on the action and consider only K\"{a}hler potential term to define the ${\cal N}=2$ supersymmetric nonlinear sigma model in two dimensions
\beq
S&=&\int d^2 \theta d^2 \bar{\theta} d^2 x K[\Phi, \Phi^\dag]\nonumber\\
&=&\int d^2 x \Bigg[g_{n \bar{m}}\left(\partial^{\mu} \varphi^n \partial_{\mu} \varphi^{* \bar{m}} +\frac{i}{2} \bar{\psi}^{\bar{m}} \sigma^{\mu}(D_{\mu} \psi)^n +\frac{i}{2} \psi^{n} \bar{\sigma}^{\mu}(D_{\mu} \bar{\psi})^{\bar{m}} +\bar{F}^{\bar{m}} F^{n}\right) \nonumber\\
&&-\frac{1}{2} K_{,nm \bar{l}} \bar{F}^{\bar{l}} \psi^n \psi^m -\frac{1}{2} K_{,n \bar{m} \bar{l}} F^{n} \bar{\psi}^{\bar{m}} \bar{\psi}^{\bar{l}}+\frac{1}{4} K_{,nm \bar{k} \bar{l}} (\bar{\psi}^{\bar{k}} \bar{\psi}^{\bar{l}})(\psi^n \psi^m)\Bigg],\label{action}
\eeq
where $\Phi^n$ denote chiral superfields, whose components fields are complex scalars $\varphi^n(x)$, Dirac fermions $\psi^n(x)$ and complex auxiliary fields $F^n(x)$. The K\"{a}hler metric of the target space $g_{i \bar{j}}$ is given by the K\"{a}hler potential
\beq
g_{i \bar{j}}(\varphi, \varphi^{\dagger})
=\frac{\partial^2 K(\varphi, \varphi^{\dagger})}{\partial \varphi^i \partial \varphi^{* \bar{j}}}.
\eeq
Higher derivative terms are not included in this paper to avoid the introduction of negative metric states.

For this nonlinear sigma model, the WRG eq.(\ref{WRG-1}) has been derived in Ref.\cite{HI}.
The scalar part of the WRG equations takes a simple form
\beq
&&\frac{d}{dt}\int d^2 x g_{i \bar{j}} (\partial_\mu \varphi)^i (\partial^\mu \varphi^*)^{\bar{j}}\nonumber\\
&&=\int d^2 x \Big[-\frac{1}{2 \pi} R_{i \bar{j}} -\gamma \Big(\varphi^k g_{i \bar{j},k}+ \varphi^{*\bar{k}}g_{i \bar{j},\bar{k}}+2g_{i \bar{j}} \Big) \Big] (\partial_\mu \varphi)^i  (\partial^\mu \varphi^*)^{\bar{j}}.\nonumber\\
\eeq

The field variables $\varphi^n(x)$ are assumed $t$ independent by a suitable rescaling of fields, which introduces the anomalous dimension $\gamma$ in return. What depends on $t$ is the infinite number of coupling constants included in the K\"{a}hler metric.  
From this WRG equation, the $\beta$ function for the K\"{a}hler metric is given by
\beq
\frac{d}{dt}g_{i \bar{j}}&=&-\frac{1}{2 \pi}R_{i \bar{j}} -\gamma \Big[\varphi^k g_{i \bar{j},k} +\varphi^{* \bar{k}}g_{i \bar{j},\bar{k}}+2g_{i \bar{j}} \Big]\nonumber\\
&\equiv&-\beta (g_{i \bar{j}}).\label{beta}
\eeq
Note that our $\beta$ function reduces to the Ricci tensor when the anomalous dimension of the fields vanishes. The second term which is proportional to the anomalous dimension $\gamma$ is not reparametrization invariant because of the renormalization condition of the fields breaks reparametrization invariance.
The fermion part also gives the same WRG equation because of the supersymmetry.
Since the K\"{a}hler metric contain the infinite number of coupling constants, the above WRG equation is an infinite set of differential equations for these coupling constants. 
In the next section, we investigate the conformal theories defined as the fixed-points of this renormalization group equation.

\section{Fixed-point of SU(N) symmetric WRG equation}\label{solution}

In this section, let us derive the action of the conformal field theory 
corresponding to the fixed-point of the $\beta$ function
\beq
\beta(g_{i \bar{j}})=\frac{1}{2 \pi}R_{i \bar{j}} +\gamma \Big[\varphi^k g_{i \bar{j},k} +\varphi^{* \bar{k}}g_{i \bar{j},\bar{k}}+2g_{i \bar{j}} \Big]
=0. \label{fpeq}
\eeq
Since Ricci curvature $R_{i\bar{j}}$ is a second derivative of the metric $g_{i \bar{j}}$, the equation is a set of coupled partial differential equations, and is very difficult to solve in general.
So we simplify the problem by assuming symmetry ${\bf U}(N)$ for the K\"{a}hler potential.
\beq
K[\varphi,\varphi^\dag]&=&\sum^{\infty}_{n=1} g_n x^n \equiv f(x)\label{SUN}
\eeq
where $x$ is the ${\bf U}(N)$ invariant combination 
\beq
x \equiv \vec{\varphi} \cdot \vec{\varphi}^\dag
\eeq
of the $N$ components scalar fields $\vec{\varphi}=(\varphi^1,\varphi^2,\cdots,\varphi^N)$. The coefficients $g_n$ play the role of an infinite number of coupling constants which depend on the cutoff scale $t$.
The K\"{a}hler potential gives the K\"{a}hler metric and Ricci tensor as follows\footnote{We use the convention: $\varphi^*_i=\delta_{i\bar{j}}\varphi^{* \bar{j}}$ and $\varphi_{\bar j}=\delta_{i\bar{j}}\varphi^{i}$.}:
\beq
g_{i \bar{j}}&\equiv&\partial_i \partial_{\bar{j}} K[\varphi,\varphi^\dag]
=f' \delta_{i \bar{j}}+f'' \varphi_i^* \varphi_{\bar{j}},\label{metric}\\
R_{i \bar{j}}&\equiv&-\partial_i \partial_{\bar{j}} \tr \ln g_{i \bar{j}}\nonumber\\
&=&-\Big[(N-1)\frac{f''}{f'} +\frac{2f''+f''' x}{f'+f'' x}  \Big]\delta_{i \bar{j}} \nonumber\\
&&-\Big[(N-1)\bigg(\frac{f^{(3)}}{f''}-\frac{(f'')^2}{(f')^2} \bigg)+\frac{3f^{(3)}+f^{(4)} x}{f'+f'' x} -\frac{(2f''+f'''x)^2}{(f'+f''x)^2} \Big]\varphi^*_{i}\varphi_{\bar{j}},\nonumber\\
\eeq
where
\beq
f'=\frac{df}{dx}.
\eeq
We substitute these metric and Ricci tensor for the $\beta$ function (\ref{beta}) and compare the coefficients of $\delta_{i \bar{j}}$ and $\varphi^i \varphi^{* \bar{j}}$ to find
\beq
\frac{\partial}{\partial t}f'&=&\frac{1}{2\pi}\Big[(N-1)\frac{f''}{f'} +\frac{2f''+f''' x}{f'+f'' x}  \Big]- 2\gamma(f'+f''x)\label{f'},\label{beta-f}\\
\frac{\partial}{\partial t}f''&=&\frac{1}{2\pi}\Big[(N-1)\bigg(\frac{f^{(3)}}{f''}-\frac{(f'')^2}{(f')^2} \bigg)+\frac{3f^{(3)}+f^{(4)} x}{f'+f'' x} -\frac{(2f''+f'''x)^2}{(f'+f''x)^2} \Big]\nonumber\\
&&-2\gamma(2f''+f'''x).\label{f''}
\eeq
Since the second equation (\ref{f''}) follows from the first equation by differentiation with respect to $x$, we discuss only the first equation.

Our differential equation (\ref{beta-f}) describes the renormalization group flow in the theory space specified by the infinite number of coupling constant in the K\"{a}hler potential.  In fact, we can derive an infinite number of coupled differential equations among the coupling constants $g_n$ by inserting (\ref{SUN}) in eq.(\ref{beta-f}).
We are specially interested in the fixed-point of eq.(\ref{beta-f}), which is 
supposed to give a scale invariant theory. The fixed-point theory is defined by the K\"{a}hler metric which satisfies the following differential equation
\beq
\frac{\partial}{\partial t}f'&=&\frac{1}{2\pi}\Big[(N-1)\frac{f''}{f'} +\frac{2f''+f''' x}{f'+f'' x}  \Big]- 2\gamma(f'+f''x)\nonumber\\
&=&0.\label{beta-0}
\eeq
To obtain the Lagrangian of the scale invariant field theory, we have to solve this differential equation.

By noting that this equation can be rewritten as
\beq
\frac{d}{dx}\Big[\ln (f')^{N-1} (f'+f''x) \Big]=4\pi \gamma \frac{d}{dx}(f'x),
\eeq
we can integrate it easily to obtain
\beq
(f')^{N-1} F'=C_1 \exp [4\pi \gamma F],\label{F}
\eeq
where
\beq
F\equiv f'x,
\eeq
and $C_1$ is an integration constant.
The normalization condition of the kinetic term,
\beq
g_{i \bar{j}}|_{x=0}=\delta_{i \bar{j}},\nonumber
\eeq
gives an initial condition
\beq
f'(0)=1, \qquad \mbox{namely}\quad F'(0)=1 \label{bc}
\eeq
which fixes $C_1=1$.

Integrating eq.(\ref{F}), we see that the solution of the differential equation satisfy the following algebraic equation:
\beq
\frac{e^{aF}}{a} \sum^{N-1}_{r=0} (-1)^r \frac{(N-1)! F^{(N-1)-r}}{(N-1-r)! a^r}=\frac{1}{N}x^N +C_2.\label{exact-sol}
\eeq
where we have introduced a constant $a \equiv -4\pi \gamma$, namely, 
we write the anomalous dimension of the scalar field using a free parameter $a$ as follows:
\beq
\gamma=-\frac{a}{4 \pi}\label{defa}.
\eeq
In NL$\sigma$M, the anomalous dimension of the scalar field can take either positive or negative value, so the parameter $a$ can also take either sign \cite{HI-anodim}.
Setting $x=0$ in eq.(\ref{exact-sol}) and using the boundary condition (\ref{bc}), we obtain
\beq
C_2=(-1)^{N-1}  \frac{(N-1)!}{a^N}.
\eeq

From this condition (\ref{exact-sol}), we can obtain the function $f'=F/x$.
Since the metric (\ref{metric}) is determined by $f'$ and $f''$, our Lagrangian is completely fixed by eq.(\ref{exact-sol}). Thus we found that the Lagrangian of the scale invariant theory has a free parameter $a$ corresponding to the anomalous dimension of the field. 

\section{Geometry of the target space of the scale invariant theory}\label{geometry}
In this section, we study the geometry of target space when the theory is scale invariant.

\subsection{One-dimensional target space}

This equation (\ref{exact-sol}) is very simple when the target manifold is of complex one-dimension.
When $N=1$, this equation reads
\beq
\frac{e^{aF}}{a}&=& x+\frac{1}{a}
\eeq
which gives 
\beq
f'=\frac{1}{ax} \ln (1+ax).
\eeq
Using this $f'$ in eq.(\ref{metric}) gives the metric of the target space
\beq
g_{i \bar{j}}=f'+f''x=\frac{1}{1+ax}.\label{CFT-metric}
\eeq
Note that the metric has only one component, and the indices $i$ and $\bar{j}$ is $1$.
The scalar curvature is given
\beq
R=\frac{a}{1+ax}.
\eeq
The property of this target manifold crucially depends on the sign of the parameter $a$.

Now, we investigate the property of the target manifold for each sign of $a$.
\begin{enumerate}
	\item When $a>0$, the anomalous dimension is negative.\\
	Since the line element is given by 
\footnote{We use $z$ for the coordinate of the manifold instead of $\varphi$ throughout this section.}
	\beq
	ds^2=\frac{|dz|^2}{1+a|z|^2}, \label{lineel}
	\eeq
	or in polar coordinate $z=re^{i\phi}$
	\beq
	ds^2=\frac{1}{1+a r^2} \Big( (dr)^2 +r^2 (d \phi)^2   \Big),\label{line-2}
	\eeq
	the volume and the distance from the origin ($r=0$) to infinity ($r=\infty$) is divergent, while the length of the circumference at the infinity is finite. Therefore, the shape of the target manifold is a semi-infinite cigar. The volume integral of the scalar curvature is also finite, giving the Euler number:
	\beq
	\chi = \frac{1}{2\pi} \int dz d\bar{z} (\det g_{i \bar{j}}) R
	=\frac{1}{2\pi} \int dz d\bar{z} \frac{1}{(1+ a|z|^2)^2}=1,
	\eeq
	which is equal to that of a disc.

\begin{figure}[h]
\begin{center}
\includegraphics[width=7cm]{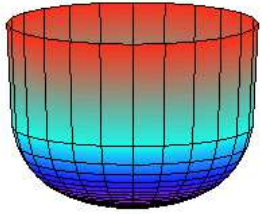}
\caption{The target manifold for $a=+1$ embedded in $3$-dimensional flat Euclidean spaces. It looks like a semi-infinite cigar with a radius $\sqrt{\frac{1}{a}}$. Our metric (\ref{embed1}) is the induced metric on this surface.}
\label{fig:Euclidean}
\end{center}
\end{figure}

	Let us embed the manifold in $3$ dimensional Euclidean spaces.
	When the hyperplane has the rotational symmetry, the line element can be written using the cylindrical coordinates
	\beq
	ds^2=(dh)^2 +(d \rho)^2 +\rho^2 (d \phi)^2.\label{line-1}
	\eeq
	Where the height is a function of the radius $\rho$, 
	\beq
	h=g(\rho).\label{h}
	\eeq
	From eq.(\ref{h}), the line element can be rewritten
	\beq
	ds^2 =(1+(g' (\rho))^2)(d \rho)^2 +\rho^2 (d\phi)^2,\label{line-3}
	\eeq
	where $g'$ is the derivative of the function $g(\rho)$ in term of $\rho$.
	Now we transform the line element for the target metric (\ref{CFT-metric}) 
	to the form of eq.(\ref{line-3}) by a change of variable
	\beq
	\rho=\frac{r}{\sqrt{1+a r^2}},
	\eeq
	which is a one-to-one mapping from the entire plane $0\le r < \infty$ to a disc $0\le \rho <\frac{1}{\sqrt{a}}$.
	Then eq.(\ref{line-2}) can be rewritten
	\beq
	ds^2 =\frac{1}{(1-a \rho^2)^2}(d \rho)^2 +\rho^2 (d \phi)^2.\label{line-4}
	\label{embed1}
	\eeq
	Comparing eq.(\ref{line-3}) with eq.(\ref{line-4}), we obtain the height function $h=g(\rho)$ as follow:
	\beq
	h&=&\int_0^{\rho} d \rho g'(\rho)
	=\int_0^{\rho} d \rho \sqrt{\frac{1}{(1-a \rho^2)^2}-1}\nonumber\\
	&=&-\frac{1}{2\sqrt{a}} \Big(\ln |\frac{\sqrt{2-a \rho^2}-1}{\sqrt{2-a \rho^2}+1}|+2\sqrt{2-a \rho^2}-\ln |\frac{\sqrt{2}-1}{\sqrt{2}+1}| -2\sqrt{2} \Big)\nonumber\\\label{h-final}.
	\eeq
	Figure \ref{fig:Euclidean} shows the manifold embedded in $3$-dimensional flat Euclidean spaces.
	The distance between any two points is measured along the shortest path on the surface in the Euclidean spaces.

	\item When $a<0$, the anomalous dimension is positive. \\
	In this case, the metric and scalar curvature read
	\beq
	g_{i \bar{j}}&=&\frac{1}{1-|a|x}\\
	R&=&\frac{-|a|}{1-|a|x}.
	\eeq
	This metric is ill-defined at the boundary $|z| \sim \frac{1}{\sqrt{-a}}$.
	It is not merely the coordinates singularity because the scalar curvature is divergent at the boundary.
	Although the volume integral is divergent, the distance to the boundary is finite.
	Now, let us embed this manifold in flat space.
	Note the eq.(\ref{h-final}) is imaginary if $a<0$.
	Thus the manifold is embedded as a space-like surface in the flat Minkowski space.
	Figure \ref{fig:Minkowski} shows the manifold embedded in the $3$-dimensional flat Minkowski space.
	
\begin{figure}[h]
\begin{center}
\includegraphics[width=7cm]{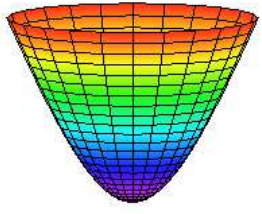}
\caption{The target manifold for $a=-1$, embedded in flat Minkowski space. The vertical axis has negative signature. In the asymptotic region $\rho \rightarrow \infty$, the surface approaches to the lightcone.}
\label{fig:Minkowski}
\end{center}
\end{figure}
	
\end{enumerate}

\subsection{Higher dimensional target spaces}

\begin{figure}
\begin{center}
\includegraphics[width=7cm]{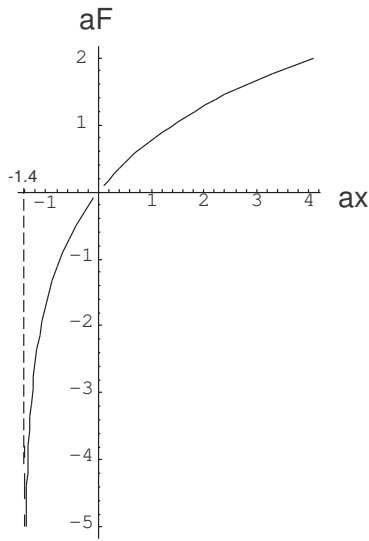}
\caption{$aF$ as a function of $ax$ for $N=2$ and $|a|=1$. 
Since $x=zz^{\dagger}$ is positive, $ax\ge 0$ corresponds to $a>0$ while 
$ax\le 0$ corresponds to $a<0$.}
\label{fig:aFax}
\end{center}
\end{figure}

Consider the conformal field theories whose target space have more than two dimensions, and investigate the property of the target manifolds.
For $N\ge 2$, we have to solve the algebraic equation (\ref{exact-sol}) 
which reads for $N=2$, for example,
\beq
e^{aF}(aF-1)=\frac{1}{2}(ax)^2 -1.\label{N=2}
\eeq
Figure \ref{fig:aFax} displays $aF$ as a function of $ax$ for $N=2$ and $|a|=1$.

K\"{a}hler potential in the neighborhood of the origin is easily obtained by solving the equation (\ref{exact-sol}) 
\beq
f(x)=x-\frac{a}{2(N+1)}x^2+\cdots .\label{weak}
\eeq
The asymptotic behaviors crucially depend on the sign of the parameter $a$, so that we will discuss them separately. 
\begin{enumerate}
	\item $a>0$ case\\
	Figure \ref{fig:aFax} show the function $f'$, which is the diagonal component of the target metric, goes to infinity as $x \rightarrow \infty$. When $aF$ goes to infinity, the $r=0$ term of eq.(\ref{exact-sol}) gives the dominant contribution in the left-hand side. To find the asymptotic behavior in this region, we retain only the dominant terms and solve
	\beq
	e^{aF} (aF)^{N-1} \approx \frac{1}{N}(ax)^N.\label{N=2}
	\eeq
	by the iteration method
	\beq
	aF&\approx&\ln \Big(\frac{(ax)^N}{N} \Big) -(N-1)\ln (aF)\nonumber\\
	&=&\ln \Big(\frac{(ax)^N}{N} \Big) -(N-1)\ln \ln \Big(\frac{(ax)^N}{N} \Big)
	+\cdots.\nonumber
	\eeq
	where we dropped other terms that vanish as $ax \rightarrow \infty$.
	Then we obtain the functions $f'$  
	\beq
	f'&\approx& \frac{1}{ax} \ln \frac{(ax)^N}{N}-\frac{N-1}{ax} \ln \ln \frac{(ax)^N}{N}+ \cdots .\label{fprime}	
	\eeq 
	The distance along the straight line in the radial direction is written 
	\beq
	g_{i \bar{j}} dz^i dz^{*\bar{j}}&=&(f' \delta_{i \bar{j}}+f'' z_{\bar{j}}z^*_{i}) dz^i dz^{*\bar{j}} \nonumber\\
	&\approx&(f'+|r|^2 f'')|dr|^2=\frac{N}{a|r|^2}|dr|^2 \label{asympto2}
	\eeq
	Here we have defined the complex radial coordinate $r$ and the angle variables by
	\beq
	z^1=u^1 r, \ \cdots\ , \ z^{N-1}=u^{N-1}r,\  z^N=r. \label{cpn}
	\eeq
	
	The asymptotic behavior (\ref{asympto2}) for any $N$ is similar to $N=1$ case,    in which the metric in the asymptotic region is given by
	\beq
	g_{1 \bar{1}}\approx \frac{1}{ax}.\label{aympto1}
	\eeq
	The asymptotic behavior of the K\"{a}hler potential (\ref{SUN}) can be found by integrating eq.(\ref{fprime}) 
	\beq
	K(zz^*) & \approx & \frac{N}{2a}(\log{(azz^*)})^2+\cdots\nonumber\\
	&=& \chi\chi^*+2{\sqrt{\frac{N}{a}}}{\frak Re}{\chi}\log{\left(1+|u^1|^2+\cdots +|u^{N-1}|^2\right)}+\cdots \label{kpcp}
	\eeq
	where we have dropped holomorphic and anti-holomorphic terms by a K\"{a}hler transformation and defined $\chi$ by
	\beq
	\chi=\sqrt{\frac{N}{a}}\log{\sqrt{a}r}.\label{fieldtr}
	\eeq
	For fixed values of the radius $r$, (\ref{kpcp}) is the K\"{a}hler potential for the Fubini-Study metric of the complex projective space $CP^{N-1}$, whose size is fixed by ${\frak Re}{\chi}$. Therefore, our target space is the direct product of the complex line represented by $\chi$ and the $CP^{N-1}$ represented by $(u^1,\cdots,u^{N-1})$ in the asymptotic region.

	\item $a<0$ case \\
	Figure \ref{fig:aFax} shows that the allowed region $ax\le 0$ is limited inside of a ball $|\vec{z}|<(\frac{\sqrt{2}}{|a|})^\frac{1}{2}$ for $N=2$ as in the case of $N=1$. By assuming $ax \sim -\sqrt{2}+\epsilon$ near the boundary,  we can reduce eq.(\ref{N=2}) to
	\beq
	e^{aF} aF \sim -\sqrt{2}\epsilon\nonumber
	\eeq 
	which can be solved by the iteration method
	\beq
		-aF &=& \ln \frac{1}{\sqrt{2}\epsilon}+\ln (-aF)\nonumber\\
	&=&\ln \frac{1}{\sqrt{2}\epsilon} +\ln\ln (\frac{1}{\sqrt{2}\epsilon}) +\cdots .\nonumber
	\eeq
	Because of $\epsilon<<1$, the behavior of the function $f'=\frac{F}{ax}$ near the boundary is given
	\beq
	f' \sim \frac{1}{ax} \ln (1+\frac{ax}{\sqrt{2}}),\label{N=2-2}
	\eeq
	which leads to the curvature singularity at the boundary.

	Similarly, the allowed region in $z$-plane for general $N$ is 
	\beq
	|\vec{z}|<(N!)^{\frac{1}{2N}}|a|^{-\frac{1}{2}}.
	\eeq
	The asymptotic behavior of the function $f'$ near the boundary 
	\[
	f'(x)\sim \frac{1}{ax}\ln (1+\frac{ax}{(N!)^{1/N}})
	\]
	leads to the curvature singularity at the boundary.
	\end{enumerate}

To summarize, we found that the target spaces of the scale invariant theory with nontrivial anomalous dimension are noncompact and well-behaved at the infinity for $a\ge 0$, while they have formidable curvature singularity at the boundary for $a<0$.

\section{Property of the field theory at the fixed-point}\label{fixedpoint}
In this section, let us discuss the property of the scale invariant field theories for $a\ge 0$.
From (\ref{weak}), the K\"{a}hler potential is given as a power series of
$\varphi\varphi^{\dagger}$
\beq
K(\vec{\varphi}\vec{\varphi}^{\dagger})=\vec{\varphi}\vec{\varphi}^{\dagger}-\frac{a}{2(N+1)}(\vec{\varphi}\vec{\varphi}^{\dagger})^2+\cdots .\label{weakth}
\eeq
All coefficients in this series are expressed by a single parameter $a$.
When $a=0$ this Lagrangian reduces to a free field theory
\[
	{\cal L}=|\partial_{\mu}\vec{\varphi}|^2 +\frac{i}{2} \bar{\vec{\psi}} \sigma^{\mu}\partial_{\mu} \vec{\psi} +\frac{i}{2} \vec{\psi} \bar{\sigma}^{\mu}\partial_{\mu} \bar{\vec{\psi}}.
\]
with the two-point function
\beq
\langle \varphi^m(x)\varphi^{*\bar{n}}(y)\rangle
=\delta^{m\bar{n}}\frac{1}{2\pi}\log{\frac{1}{|x-y|}}, \label{freeprop}
\eeq
which corresponds to a vanishing anomalous dimension $\gamma=0$. 
Higher order terms of $a$ in (\ref{weakth}) introduce the interaction, 
which gives the nontrivial anomalous dimension given by (\ref{defa}) while 
$\beta$ function remains zero by the contributions of further higher order terms in (\ref{weakth}). 
Although the Lagrangian (\ref{weakth}) gives a good description for small values of the fields $\varphi(x)$, we have to use another Lagrangian derived from the K\"{a}hler potential (\ref{kpcp}) to describe the phenomena for large values of the fields. Equations (\ref{kpcp}) and (\ref{fieldtr}) indicate that ${\frak Im}{\chi}$ is a periodic variable with a period $2\pi\sqrt{\frac{N}{a}}$. 
The K\"{a}hler potential for the Fubini-Study metric (\ref{kpcp}) implies that the squared radius of the $CP^{N-1}$ is proportional to $2\sqrt{\frac{N}{a}}{\frak Re}{\chi}$. 

In the presence of the nontrivial anomalous dimension $\gamma$, the two-point function has to behave 
\beq
\langle \varphi^m(x)\varphi^{*\bar{n}}(y)\rangle
\propto \delta^{m\bar{n}}\frac{1}{|x-y|^{2\gamma}}\label{anprop}
\eeq
because of the scale invariance.
Although it is difficult to solve field theory with interaction, 
we can obtain this kind of behavior in the strong coupling limit $a\rightarrow \infty$ of $N=1$ model defined by the metric (\ref{CFT-metric}). Since the target space approaches to a cylinder at the infinity as is shown in the figure \ref{fig:Euclidean}, it also approaches to another free field theory when $a$ goes to infinity. In fact, the bosonic part of the Lagrangian for $N=1$ model
	\beq
	{\cal L}=\frac{1}{1+a|\varphi|^2}|\partial_{\mu}\varphi|^2\label{N1lag}
	\eeq
	reduces to that of a free field theory when $a\rightarrow \infty$
	\[
	{\cal L}=|\partial_{\mu}\chi|^2,
	\]
	where
	\beq
	\chi=\frac{1}{\sqrt{a}}\log{(\sqrt{a}\varphi)}.\label{phichi}
	\eeq
Using this relation (\ref{phichi}), we can evaluate the two-point function of $\varphi$ by using the free propagator (\ref{freeprop}) for $\chi$
\beq
\langle \varphi(x)\varphi^{*}(y)\rangle&=&
\frac{1}{a}\langle e^{\sqrt{a}\chi(x)}e^{\sqrt{a}\chi^{*}(y)}\rangle\nonumber\\
&=&\frac{1}{a}\exp{(a\langle\chi(x)\chi^*(y)\rangle)}
\propto \frac{1}{|x-y|^{\frac{a}{2\pi}}}.\label{intpro}
\eeq
 Thus, we find the anomalous dimension of opposite sign 
in the strong coupling regime. This seems to be an indication 
that the anomalous dimension aquires higher order corrections 
in the strong coupling regime. Study of the strong coupling 
expansion to reconcile this discrepancy is left for future works. 

Although the real dimension of the target manifold is two in $N=1$ model, it looks like a cylinder with radius $\sqrt{\frac{1}{a}}$ when viewed from far away places.

\section{Conclusion}
In order to find non-trivial conformal field theories with anomalous dimension, we used the WRG equation. In solving the WRG equation, we have assumed ${\bf U}(N)$ symmetry to reduce the coupled partial differential equation to an ordinary differential equation. The new class of conformal field theories have one parameter $a$, corresponding to the anomalous dimension of the scalar field. The novel conformal field theories are well behaved for positive $a$, while they have curvature singularities at the boundary for $a<0$. We obtained the Lagrangian explicitly for $N=1$, which allows both the strong and weak coupling expansion. The target space in this case looks like a semi-infinite cigar with one-dimension compactified to a circle. It is interesting to examine the conformal symmetry of our new models both in the weak and in the strong coupling expansion.

After the completion of this work we came to know that our metric 
was also discussed in other context. 
The $N=1$ model was proposed by Witten as a model of 
the two dimensional black hole~\cite{2DBH}, which was subsequently 
generalized by Kiritsis, Kounnas and Lust as consistent 
backgrounds of the superstrings in the presence of the 
dilaton~\cite{KKL}. Hori and Kapustin proposed to quantize 
these models by means of linear sigma models with massive 
gauge field in the Stueckelberg formalism~\cite{HK}.

\section{Acknowledgements}
This work was supported in part by the Grant-in-Aid for Scientific
Research (\#13640283 and \#13135215). We would like to thank 
Toshio Nakatsu and Anton Kapustin for pointing out 
references~\cite{2DBH, KKL, HK}. 


\end{document}